\documentclass[showpacs,preprintnumbers,amsmath,amssymb,twocolumn,superscriptaddress,prl]{revtex4}
\usepackage{graphicx}
\usepackage{amsfonts}
\usepackage{amsmath, amsthm, amssymb}
\usepackage{dsfont}

\begin{document}
\title[Effective theory of fluctuating orbital currents in high-$T_c$ cuprates]
{Effective theory of fluctuating orbital currents in high-$T_c$ cuprates}
\author{Kjetil B{\o}rkje}
\affiliation{Department of Physics, Norwegian University of
Science and Technology, N-7491 Trondheim, Norway}
\author{Asle Sudb{\o}}
\affiliation{Department of Physics, Norwegian University of
Science and Technology, N-7491 Trondheim, Norway}
\date{Received \today}
\begin{abstract}
  We derive an effective dissipative quantum field theory for
  fluctuating orbital currents in clean $CuO_2$ sheets of high-$T_c$
  cuprates, based on a three-band model.  The Coulomb repulsion term
  between $Cu$- and $O$-sites is decoupled in terms of current operators
  representing horizontal and vertical parts of circulating currents
  within each $CuO_2$ unit cell of the lattice. The model has ordering
  of currents at finite temperatures. The dissipative kernel in
  the model is of the form $|\omega|/|{\bf q}|$, indicating Landau
  damping. Applications of the effective theory to other models are 
  also discussed.

\end{abstract}
\maketitle
Constructing an effective description of the long-wavelength and
low-energy physics of high-$T_c$ superconducting cuprates represents a
profound and formidable problem in physics. Such a description must be consistent
with experimental observations of several anomalous normal state
properties of these systems. Varma has recently proposed that quantum
critical fluctuations associated with the breakup of a subtle order,
involving circulating currents, could induce the observed anomalous normal
state properties of high-$T_c$ superconductors
\cite{Varma}. Essentially, the associated quantum critical
fluctuations are suggested to produce a fluctuation spectrum resulting
in a Marginal Fermi Liquid \cite{MFL_Varma}. Recently, such a spectrum
has been derived from a conjectured effective field theory of
circulating currents \cite{Aji_Varma}. It should, however, be mentioned
that a recent numerical evaluation of the current-current correlations
in a three-band $t-J$-model with $24$ sites, where doubly occupied sites have been 
projected out, shows no evidence of the orbital current pattern proposed by 
Varma \cite{Greiter_Thomale}. 
\par
\begin{figure}[htbp]
  \centerline{\hbox{\includegraphics[width=45mm]{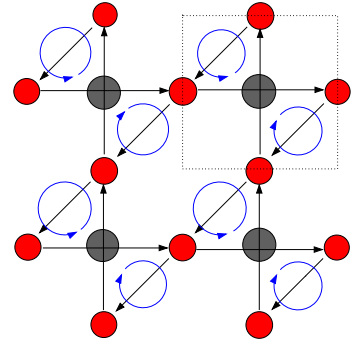}}}
  \caption{(Color online) The circulating current phase $\Theta_{II}$ 
  \cite{Varma}. $Cu$ sites are grey circles, $O$ sites are red.  The unit 
  cell is shown by the dashed square. A staggered magnetic moment pattern 
  within each unit cell that repeats from unit cell to unit cell (the curl 
  of the blue directed circles) is indicated. }
    \label{orbital_currents}
\end{figure}

The particular form of proposed  order involves circulating currents 
within a $CuO_2$ unit cell where the currents run horizontally and vertically
through a $Cu$ site and close by {\it direct hopping between $O$ orbitals},
as in Fig. \ref{orbital_currents}. Three other equivalent
patterns may be found by reversing the direction of the current through 
each $Cu$-site in the horizontal and vertical directions. 
This results in a pattern of {\it staggered} orbital magnetic moments within each 
unit cell, such that the pattern repeats from unit cell to unit cell.  A magnetic 
intensity of the type associated with the above orbital magnetism has recently been 
reported \cite{Fauque_Mook}. Since no obvious thermodynamic singularities have so far 
been reported at the pseudogap-line in the cuprates, it is important to investigate 
whether or not the proposed models for this novel type of order imply the  
presence or absence of prominent signals in such quantities as specific heat or 
(indirectly) magnetization. Other staggered orbital magnetic patterns have also 
been proposed, most notably the extension of the staggered flux phase \cite{AM1988} 
to finite doping \cite{DDW}. We emphasize that our justification for focusing on the orbital
current pattern proposed by Varma \cite{Varma}, are the experiments reported in 
Ref. \onlinecite{Fauque_Mook}. 
\par
We derive an effective quantum field theory for fluctuating orbital
currents from a microscopic description of clean $CuO_2$ planes. We
are primarily interested in investigating the intrinsic effects such
fluctuations have on the physics of the cuprates. We therefore neglect
disorder, as was also done in Ref. \onlinecite{Aji_Varma}.  With ever
improving sample quality, we expect that the effective theory we
derive should be useful. The starting point is the three-band model 
$H
= \sum_{{\bf r}, \sigma} \varepsilon_d d^{\dagger}_{{\bf r}, \sigma}
d_{{\bf r},\sigma} + K_{pd} + K_{pp} + H^{(1)}_{\rm{int}} +
H^{(2)}_{\rm{int}}$, where $K_{pd} = t_{pd} \sum_{{\bf r}, \sigma} [
d^{\dagger}_{{\bf r}, \sigma} (p_{x,{\bf r} + \frac{a}{2}
  {\bf\hat{x}}, \sigma} - p_{x,{\bf r} - \frac{a}{2} {\bf \hat{x}},
  \sigma} - p_{y, {\bf r} + \frac{a}{2} {\bf \hat{y}}, \sigma} + p_{y,
  {\bf r} - \frac{a}{2} {\bf \hat{y}}, \sigma}) + \text{h.c.} ]$,
$K_{pp} = - t_{pp} \sum_{{\bf r}, \sigma} [ ( p^\dagger_{x,{\bf r} +
  \frac{a}{2} {\bf\hat{x}}, \sigma} - p^\dagger_{x,{\bf r} -
  \frac{a}{2} {\bf \hat{x}}, \sigma} ) (p_{y,{\bf r} + \frac{a}{2}
  {\bf\hat{y}}, \sigma} - p_{y,{\bf r} - \frac{a}{2} {\bf \hat{y}},
  \sigma} ) + \text{h.c.} ]$ and $H^{(2)}_{\rm{int}} = V \sum_{{\bf
    r}, \sigma, \sigma'} n_{d,{\bf r}, \sigma} ( n_{p_x, {\bf r} +
  \frac{a}{2} {\bf \hat{x}},\sigma'} + n_{p_x, {\bf r} - \frac{a}{2}
  {\bf \hat{x}},\sigma'} + n_{p_y, {\bf r} + \frac{a}{2} {\bf
    \hat{y}},\sigma'} + n_{p_y, {\bf r} - \frac{a}{2} {\bf
    \hat{y}},\sigma'} )$. We work with electron operators and the
vacuum is defined as empty $d_{x^2-y^2}$, $p_x$ and
$p_y$ orbitals. The ${\bf r}$-sum runs over the $Cu$-lattice. The
$Cu$-$O$ and $O$-$O$ hopping is governed by the parameters $t_{pd}$
and $t_{pp}$, respectively, whereas $\varepsilon_d$ is the difference
in on-site energy between the copper and oxygen orbitals. The term
$H^{(1)}_{\rm{int}}$ represents on-site repulsion terms, for which we
make the crude assumption that their effect is to merely renormalize
the hopping parameters $t_{pd} \to \bar{t}_{pd} = |x| t_{pd}$, $t_{pp}
\to \bar{t}_{pp} = |x| t_{pp}$, where $|x|$ is the deviation from
half-filling \cite{Varma}.  We also assume the $O$-$O$ repulsion to be
small. Hence, we only consider explicitly $H^{(2)}_{\rm{int}}$, the
$Cu$-$O$-repulsion.
\par
The interaction-term $H^{(2)}_{\rm{int}}$ can be decoupled \cite{HS} in
terms of bosonic fields coupling to the bilinear fermion operators
$A^{(i)}_{{\bf q}, \sigma, \sigma'} \equiv N^{-1/2}
\sum_{\bf k} \big( a_{x,{\bf k - q}}^{(i)} \, p^\dagger_{x,{\bf k -
    q},\sigma'} + a_{y,{\bf k - q}}^{(i)} \, p^\dagger_{y,{\bf k -
    q},\sigma'} \big) d_{{\bf k},\sigma}$ with $i=1,..,4$
\cite{Varma}. Here, $N$ is the number of $Cu$ lattice sites. We define
$a_{x,{\bf k}}^{(1)} = a_{x,{\bf k}}^{(2)} = \sin(k_x a/2) \equiv
s_{x,k}$, $a_{x,{\bf k}}^{(3)} = a_{x,{\bf k}}^{(4)} = \cos(k_x a/2)
\equiv c_{x,k}$ and $a_{y,{\bf k}}^{(1)} = -a_{y,{\bf k}}^{(2)} =
s_{y,k}$, $a_{y,{\bf k}}^{(3)} = -a_{y,{\bf k}}^{(4)} = c_{y,k}$,
where $a$ is the $Cu$-$Cu$ lattice constant. A discussion of $\langle
A^{(i)}_{0, \sigma, \sigma'} \rangle$ as translational invariant
order parameters in the cuprates is found in \cite{Varma}. While $i=2$
transforms as the kinetic energy, $i=1$ and $i=3,4$ {\it give rise to
  different current patterns}. Since the observed magnetic signal
\cite{Fauque_Mook} is consistent with the current patterns of $i=3,4$,
we keep only this in what follows. An effective model for the
$i=1$-part was considered in \cite{Lee_Choi}. Observe the relation
$N^{-1/2} \sum_{\bf k} c_{x,k-q} p^\dagger_{x,{\bf k-q},\sigma'} d_{{\bf
    k},\sigma} = 1/4 \left(\kappa^x_{{\bf q},\sigma,\sigma'} +
  i j^x_{{\bf q},\sigma,\sigma'}\right)$, where, in real space,
\begin{equation}
  j^x_{{\bf r},\sigma,\sigma'} \equiv \frac{i}{2}
\left[ d^\dagger_{{\bf r},\sigma} \left(p_{x,{\bf r} + \frac{a}{2}
    {\bf\hat{x}}, \sigma'} + p_{x,{\bf r} - \frac{a}{2}
    {\bf\hat{x}}, \sigma'}\right) - \text{h.c.} \right].
\end{equation}
In a unit cell centered on $Cu$, this is proportional to the current
from the left oxygen to the copper {\it plus} the current from copper
to the right oxygen. We define $j^y_{{\bf r},\sigma,\sigma'}$ in the
same way, but with a minus sign due to the $d$-wave symmetry of the
$Cu$-orbital. Finite expectation values of $\kappa^{x(y)}_{{\bf
    r},\sigma,\sigma'}$ would correspond to Landau-Pomeranchuk
instabilites, believed not to be relevant in the cuprates. Thus, we
retain only the decoupling fields that correspond to spin diagonal
expectation values of the operators $j^{x(y)}_{{\bf
    r},\sigma,\sigma'}(\tau)$, since $\langle j^{x(y)}_{{\bf
    r},\sigma,\sigma}(\tau) \rangle \neq 0$ in the current pattern
depicted in Fig. \ref{orbital_currents}. The fields retained,
$J^{x(y)}_{\bf r}(\tau)$, are real and $\langle J^{x(y)}_{{\bf
    r}}(\tau) \rangle = V \langle j^{x(y)}_{{\bf
    r},\sigma,\sigma'}(\tau) \rangle \delta_{\sigma,\sigma'}$, {\it
  i.e.} the fields represent charge currents on horizontal and
vertical $O$-$Cu$-$O$-links.  The fields $J^{x(y)}_{\bf r}(\tau)$ and
the fermions are coupled by particle-hole excitations of the form $ i
\sum_{{\bf k},{\bf q},\sigma} \left(J^x_{-{\bf q}} c_{x,k-q} \,
  p^{\dagger}_{x,{{\bf k-q}},\sigma} d_{{\bf k},\sigma} - (x \to y) -
  \text{h.c.}  \right)$, where the time dependence was omitted. It is
important to keep in mind that the bosonic fields $J^{x(y)}_{\bf
  r}(\tau)$ transform as vectors under a change of coordinate
system. Note that we could also have chosen the arguments of the $a^{(i)}$'s to be
${\bf k}$ and not ${\bf k-q}$ in $A^{(i)}_{{\bf q}, \sigma, \sigma'}$,
corresponding to a decoupling in terms of currents defined on
horizontal and vertical $Cu$-$O$-$Cu$-links.
\par
Integrating out the fermion fields, we obtain
the partition function as $Z = \int DJ^{x} D J^{y} ~ e^{-S}$, where
the effective action is given by $S = \frac{1}{2 V} \sum_{{\bf q},
  \omega_\nu} \left[J^{x}_{{\bf q}}(i \omega_\nu) J^{x}_{-{\bf q}}(- i
  \omega_\nu) + J^{y}_{{\bf q}}(i \omega_\nu) J^{y}_{-{\bf q}}(- i
  \omega_\nu) \right] - \text{Tr} \ln \left[ {\cal G}_0^{-1} + \Sigma
\right]$. Using the gauge transformation $p_{x,{\bf k},\sigma} \to i
p_{x,{\bf k},\sigma}, \ p_{y,{\bf k},\sigma} \to -i p_{y,{\bf
    k},\sigma}$, we have
\begin{widetext}
\begin{eqnarray}
{\cal G}^{-1}_{0, {\bf k}_1 {\bf k}_2,\sigma_1\sigma_2} (i \omega_{n_1}, i \omega_{n_2}) =
\delta_{ {\bf k}_1,{\bf k}_2 }  \delta_{n_1,n_2} \delta_{\sigma_1,\sigma_2}
\left(
\begin{array}{ccc}
-i \omega_{n_1} + \varepsilon_d - \mu & 2 t_{pd} s_{x,k_1} & 2 t_{pd} s_{y,k_1} \\
 2 t_{pd} s_{x,k_1} & -i \omega_{n_1} - \mu  & 4 t_{pp} s_{x,k_1} s_{y,k_1} \\
 2 t_{pd} s_{y,k_1} &  4 t_{pp} s_{x,k_1} s_{y,k_1} &   -i \omega_{n_1} - \mu
\end{array}
\right),
\end{eqnarray} 
\begin{eqnarray}
  \Sigma_{{\bf k}_1  {\bf k}_2,\sigma_1\sigma_2} (i \omega_{n_1}, i \omega_{n_2}) =
  \frac{\delta_{\sigma_1,\sigma_2}}{\sqrt{\beta N}} \left(
\begin{array}{ccc}
0 & c_{x,k_2} J^{x}_{{\bf k}_{12}} (i \omega_{1 2})   &  c_{y,k_2} J^{y}_{{\bf k}_{12}} (i \omega_{12}) \\
    c_{x,k_1} J^{x}_{{\bf k}_{12}} (i \omega_{1 2})   &   0           &                             0  \\
    c_{y,k_1} J^{y}_{{\bf k}_{12}} (i \omega_{1 2})   &   0           &                             0
\end{array}
\right),
\end{eqnarray} 
\end{widetext}
where we have defined ${\bf k}_{12} \equiv {\bf k}_1 - {\bf k}_2$ and
$\omega_{12} \equiv \omega_{n_1} - \omega_{n_2}$. For $t_{pp} = 0$,
the non-interacting part of the problem ${\cal G}_0^{-1}$ may easily
be diagonalized into three quasiparticle bands $E^{(0)}_{\bf k}=0,
E^{(\pm)}_{\bf k} = \varepsilon_d/2 \pm \sqrt{(\varepsilon_d/2)^2 + 4
  t_{pd}^2 (s_{x,k}^2 + s_{y,k}^2)}$, of which $E^{(0)}_{\bf
  k},E^{(-)}_{\bf k}$ are full and $E^{(+)}_{\bf k}$ is partially
filled.  This picture is not qualitatively altered by $t_{pp} \neq
0$. A nonzero value of $t_{pp}$ is however vital for the realization of
the current pattern. It is implicit that $\langle J^{x(y)}_{\bf r}(\tau) \rangle
\to 0$ when $t_{pp} \to 0$ \cite{Varma}. 
\par
Expanding the last term \cite{footnote_Belitz}, odd powers of $J$ vanish, such that
$\text{Tr} \ln \left[ {\cal G}_0^{-1} + \Sigma \right] = \text{Tr} \ln
{\cal G}_0^{-1} - \frac{1}{2} \text{Tr} \left[{\cal G}_0 \Sigma
\right]^2 + {\cal O}(J^4)$, where $-\text{Tr} \ln {\cal G}_0^{-1}$
gives the free energy of the non-interacting system, and $\Sigma$
involves the fluctuating fields $J^{x}$ and $J^{y}$. To second order in
the fields $J^{x(y)}$ and in space and imaginary time gradients, we
have derived a quantum dissipative effective action $S = S_{\rm{C}} +
S_{\rm{Q}}$, where
\begin{eqnarray}
S_{\rm{C}} & = & \sum_{{\bf q},\omega_\nu} \sum_{i,j=x,y}
G^{-1}_{\text{C},ij} \, J^i_{{\bf q}}(i \omega_\nu) J^j_{{\bf -q}}(-i \omega_\nu) ,
 \nonumber \\
S_{\rm{Q}} & = &  \sum_{{\bf q},\omega_{\nu}} 
\sum_{i,j=x,y} G^{-1}_{\text{Q},ij} \, J^i_{{\bf q}}(i \omega_\nu)J^j_{{\bf -q}}(-i \omega_\nu) ,
\label{eff_theory}
\end{eqnarray}
with $G^{-1}_{\text{C},xx} = \alpha_c + \alpha_{l} \, q_x^2 +
\alpha_{t} \, q_y^2$, $G^{-1}_{\text{C},yy} = \alpha_c + \alpha_{l} \,
q_y^2 + \alpha_{t} \, q_x^2$, $G^{-1}_{\text{C},xy} =
G^{-1}_{\text{C},yx} = \alpha_{xy} \, q_x q_y$, $G^{-1}_{\text{Q},xx}
= \alpha_0 \, \omega_\nu^2 + \alpha_{d} \frac{|\omega_\nu|}{|{\bf q}|}
\hat{q}_y^2$, $G^{-1}_{\text{Q},yy} = \alpha_0 \, \omega_\nu^2 +
\alpha_{d} \frac{|\omega_\nu|}{|{\bf q}|} \hat{q}_x^2$ and
$G^{-1}_{\text{Q},xy} = G^{-1}_{\text{Q},yx} = -\alpha_{d}
\frac{|\omega_\nu|}{|{\bf q}|} \hat{q}_x \hat{q}_y$.  Here, $\hat{q}_x
= q_x/|{\bf q}|$. The dissipation kernel is valid for $
|\omega_\nu|/|{\bf q}| \ll 1 $.  The limit $|\omega_\nu|/|{\bf q}| \gg
1$ does not contribute to dissipation. The explicit expressions for
the coefficients $\alpha_i$ are unwieldy and of limited use. The
equality of the diagonal and off-diagonal dissipation coefficients is
only correct when $t_{pp} = 0$. Changes when $t_{pp} \neq 0$ are small
and unimportant, and are neglected in the following. Note also that
this theory might not be applicable to the ordered phase, since the
Fermi surface is proposed to be gapped there \cite{Varma}. However, it
is the fluctuation spectrum in the disordered phase which is important
in connection with the Marginal Fermi Liquid hypothesis
\cite{MFL_Varma}.
\par
We have divided the action 
into a classical (C) and a quantum (Q) part. At finite temperatures, only the classical 
piece of the action $S_{\rm{C}}$ needs to be considered as far as critical properties are 
concerned. The excitation energies of the eigenmodes of $S_{\rm{C}}$ are given by
$\lambda_{\pm} = \alpha_c + (\alpha_{l} + \alpha_{t})\, q^2/2  \pm 
\sqrt{(\Delta \alpha)^2 q^4 + \gamma q_x^2 q_y^2}$, where 
$\Delta \alpha =(\alpha_{l} - \alpha_{t})/2$, and
$\gamma = \alpha_{xy}^2-(2 \Delta \alpha)^2$. Hence, for $(\alpha_{l},\alpha_{t}) > 0$, 
a uniformly ordered state is stable in the classical domain below some critical temperature, 
provided $\alpha_{xy}^2 < \alpha_{l}^2 + \alpha_{t}^2$. 
\par
The dissipation kernel essentially gives Landau damping, albeit
anisotropic due to the directional nature of the fields. The
dissipation is a result of coupling to the gapless particle-hole
excitations in the band $E^{(+)}_{\bf k}$.  The singular form
$|\omega_\nu|/|{\bf q}|$ is correct only if the order in the
horizontal and vertical currents are {\it uniform}
and not modulated at some nonzero reciprocal vector. It implies that
the dynamical critical exponent $z=3$
\cite{footnote_2}. See however Ref. \onlinecite{footnote_Belitz}.
\par
Current amplitude fluctuations are expected to be high-energy
excitations \cite{Varma} and will therefore not determine the critical
properties of the model.  Thus, we treat the fields $J^{x}_{\bf
  r}(\tau)$ and $J^{y}_{\bf r}(\tau)$ as {\it Ising
  variables}. Reverting to a real space $Cu$-lattice formulation and
setting $a = 1$, we obtain (up to constant terms)
\begin{widetext}
\begin{eqnarray}
\label{CIM_1}
S_{\rm{C}} & = &   -\int_0^{\beta} d \tau  
\left[ \sum_{\langle {\bf r}, {\bf r'}\rangle} \left(\tilde{\alpha}^x_{{\bf r},{\bf r'}} J^{x}_{\bf r}(\tau) J^{x}_{\bf r'}(\tau) 
    + \tilde{\alpha}^y_{{\bf r},{\bf r'}} J^{y}_{\bf r}(\tau) J^{y}_{\bf r'}(\tau)  \right) 
  +  \sum_{\langle \langle {\bf r} ,{\bf r'} \rangle \rangle} 
  \tilde{\alpha}^{xy}_{{\bf r},{\bf r'}} \Big( J^x_{\bf r}(\tau) J^y_{\bf
      r'}(\tau) + J^y_{\bf r}(\tau) J^x_{\bf
      r'}(\tau) \Big)
  \right], \\
S_{\rm{Q}} & = & 
\tilde{\alpha}_0 \int_0^{\beta} d \tau \sum_{\bf r}
\left[
 \left( \frac{\partial J^{x}_{\bf r}}{\partial \tau} \right)^2  + \left( \frac{\partial J^{y}_{\bf r}}{\partial \tau} \right)^2
\right] + \tilde{\alpha}_{d}  
\int_0^{\beta} d \tau \, d \tau' \sum_{{\bf r},{\bf r'}}
\sum_{i,j} \Big(J^i_{\bf r}(\tau) -  J^i_{\bf
    r'}(\tau')\Big)\mathds{K}^{ij}_{{\bf r-r'}}(\tau-\tau')
\left(J^j_{\bf r}(\tau) - J^j_{\bf r'}(\tau')\right). \nonumber 
\end{eqnarray}
\end{widetext}
\par
Here, $\langle {\bf r},{\bf r'} \rangle$ and $\langle \langle {\bf r},{\bf r'} \rangle \rangle$ denote 
nearest-neighbor and next-nearest-neighbor summations,
respectively. For ${\bf r - r'} = \pm {\bf \hat{x}}$,
$\tilde{\alpha}^x_{{\bf r},{\bf r'}} = \tilde{\alpha}_l$ and
$\tilde{\alpha}^y_{{\bf r},{\bf r'}} = \tilde{\alpha}_t$, whereas
when ${\bf r - r'} = \pm {\bf \hat{y}}$,
$\tilde{\alpha}^x_{{\bf r},{\bf r'}} = \tilde{\alpha}_t$ and
$\tilde{\alpha}^y_{{\bf r},{\bf r'}} = \tilde{\alpha}_l$. 
The parameter $\tilde{\alpha}^{xy}_{{\bf r},{\bf r'}} =
\tilde{\alpha}_{xy}$ when ${\bf r - r'} = \pm ({\bf \hat{x}} + {\bf \hat{y}})$
and $\tilde{\alpha}^{xy}_{{\bf r},{\bf r'}} =
- \tilde{\alpha}_{xy}$ when ${\bf r - r'} = \pm ({\bf \hat{x}} - {\bf
  \hat{y}})$. The coefficient $\tilde{\alpha}_d > 0$ and the positive semidefinite matrix $\mathds{K}_{{\bf
    r-r'}}(\tau-\tau') = K_{{\bf
    r-r'}}(\tau-\tau') \, {\bf \hat{g}}_{{\bf r - r'}}  \otimes
{\bf \hat{g}}_{{\bf r - r'}}$, where ${\bf \hat{g}}_{{\bf r - r'}} = ({\bf r-r'})/|{\bf r}-{\bf r}'|$ and 
$K_{\bf r}(\tau) = 1/(|{\bf r}|\sin^2( \pi \tau/\beta))$.
Fluctuations $(J^x_{\bf r} \to - J^x_{\bf r}, J^y_{\bf r} \to  J^y_{\bf r})$ corresponds to going from the depicted 
current pattern (Fig. \ref{orbital_currents}) to a new one which is obtained by a counterclockwise rotation by 
$\pi/2$, $(J^x_{\bf r} \to J^x_{\bf r}, J^y_{\bf r} \to  -J^y_{\bf r})$ corresponds to clockwise rotation of 
$\pi/2$, and $(J^x_{\bf r} \to -J^x_{\bf r}, J^y_{\bf r} \to  -J^y_{\bf r})$ to a rotation of $\pi$. It is implied 
that in the dissipation kernel, we must use a short-distance cutoff in $(\tau,{\bf r})$-space, since the expressions 
are derived in the limit of low $(\omega,{\bf q})$. The tildes on the coefficients indicate that the
model in Eq. \ref{CIM_1} is regularised on a lattice, and that the fields have been normalised to Ising-variables.
Moreover, there will be higher order (quartic) terms generated that simply involve local squares of Ising variables
multiplied by som bilinear combinatoin of Ising variables, and these will also contribute to
the coefficients of the quadratic terms even before a renormalization group analysis is carried out. These terms
are also taken into account by the tilde.  
\par
 In general, we have $\tilde{\alpha}_{l} \neq
 \tilde{\alpha}_{t}$. A current living on a
 horizontal $O$-$Cu$-$O$-link, $J^x_{\bf r}$, couples to $J^x_{{\bf r}\pm {\bf
     \hat{x}}}$ through $\tilde{\alpha}_{l}$, and to $J^x_{{\bf r} \pm
   {\bf \hat{y}}}$ through $\tilde{\alpha}_{t}$. As seen from Figure
 \ref{orbital_currents}, there is no reason for these couplings to be similar, and 
 in fact a detailed derivation shows that they are not \cite{footnote_anisotropy}.
\par
At finite temperature, we may ignore the inertial and dissipative terms, which reduces 
the model to a classical model of two coupled Ising fields. Such a classical model will 
suffice to study the breakup of the current pattern at finite temperatures, while its 
quantum critical version can only be accessed via the full dissipative field theory. Note 
also that the dissipation kernel is {\it non-local} both in imaginary time and in space. 
The latter distinguishes this dissipation term from the Caldeira-Leggett type of dissipation 
appropriate for an array of Josephson junctions \cite{Caldeira_Leggett,Chakravarty_2005}. 
The non-locality in ${\bf r}$-space is anisotropic for the same reason as for the 
nearest-neighbor coupling.
\par
Eq. (\ref{CIM_1}) may be rewritten on the form 
\begin{widetext}
\begin{eqnarray}
\label{4-state}
S_{\rm{C}}  & = &   
- \int_0^{\beta}   d \tau 
 \left\{
 \sum_{\langle {\bf r},{\bf r'} \rangle}  \Big[\bar{\alpha} \, \cos(\theta_{{\bf r},\tau}-\theta_{{\bf r'},\tau})  +
(\Delta \tilde{\alpha})_{{\bf r},{\bf r'}} \, \sin(\theta_{{\bf r},\tau}+\theta_{{\bf r'},\tau}) \Big]
   +   2 \sum_{\langle \langle {\bf r},{\bf r'} \rangle \rangle}
   \tilde{\alpha}^{xy}_{{\bf r},{\bf r'}} \, \cos(\theta_{{\bf r},\tau}+\theta_{{\bf r'},\tau}) 
\right\},  \\
 S_{\rm{Q}} & = & 2 \tilde{\alpha}_0 
\int_0^{\beta} d \tau 
  \sum_{\bf r} 
 \left( 
 \frac{\partial \theta_{{\bf r},\tau}}{\partial \tau}
 \right)^2 +
\tilde{\alpha}_{d} \int_0^{\beta} d \tau  \, d \tau' \sum_{{\bf
    r},{\bf r'}} \sum_{i,j}
 \Big(J^i(\theta_{{\bf r},\tau}) - J^i(\theta_{{\bf r'},\tau'})\Big)
 \mathds{K}^{ij}_{{\bf r-r'}}(\tau-\tau') \Big(J^j(\theta_{{\bf r},\tau}) - J^j(\theta_{{\bf r'},\tau'})\Big), \nonumber
\end{eqnarray} 
\end{widetext}
where we have used the parametrization
$\cos(\theta_{{\bf r},\tau}) = (J^x_{\bf r}(\tau)+J^y_{\bf r}(\tau))/2$, $\sin(\theta_{{\bf r},\tau}) = (J^x_{\bf r}(\tau)-J^y_{\bf r}(\tau))/2$,
and $\theta_{{\bf r},\tau} \in (0,\pi/2,\pi,3 \pi/2)$. We have defined
$\bar{\alpha} = (\tilde{\alpha}_l + \tilde{\alpha}_t)$, $(\Delta
\tilde{\alpha})_{{\bf r},{\bf r'}} = (\tilde{\alpha}_l -
\tilde{\alpha}_t)$ for ${\bf r -r'} = \pm {\bf
  \hat{x}}$ and $(\Delta \tilde{\alpha})_{{\bf r},{\bf r'}} = -(\tilde{\alpha}_l - \tilde{\alpha}_t)$
for ${\bf r -r'} = \pm {\bf \hat{y}}$.
\par
Eqs. (\ref{CIM_1}) and (\ref{4-state}) are the main results of this paper. These models 
describe a phase transition  from a disordered bosonic state (a Fermi liquid), into a 
state with bosonic order in the form of ordered orbital currents. 
\par
We next proceed to discuss some qualitative aspects. Consider first this model at 
finite temperature, where we may use the approximation $S \approx S_{\rm{C}}$. 
When $(\tilde{\alpha}_{l},\tilde{\alpha}_{t}) > 0$ and $\tilde{\alpha}_{xy}=0$, 
the current pattern in Fig. \ref{orbital_currents} repeats uniformly from unit cell 
to unit cell throughout the system in the ordered state. The  specific heat has a 
logarithmic singularity at a critical temperature determined by the condition 
$\sinh(2 \beta_c \tilde{\alpha}_{l}) \sinh(2 \beta_c \tilde{\alpha}_{t}) = 1$, 
where $\beta_c =1/T_c$.  Anisotropy in the nearest-neighbor couplings suppresses the 
critical temperature and critical amplitudes, and narrows the critical region, but 
does not alter the universality class of the phase transition \cite{Onsager}.  When 
$(\tilde{\alpha}_{l}, \tilde{\alpha}_{t}) = 0$ and $\tilde{\alpha}_{xy} \neq 0$, the 
ground state of the system features a striped phase in the diagonal directions, 
irrespective of the sign of $\tilde{\alpha}_{xy}$. Note  also that when 
$\text{sign}(\tilde{\alpha}_{l}) \neq \text{sign}(\tilde{\alpha}_{t})$ and 
$\tilde{\alpha}_{xy} = 0$, one obtains order with a period of twice the lattice 
constant. 
\par
The dissipative term in this model comes from the coupling of the
bosonic current fields to particle-hole excitations in the partially
filled band $E^{(+)}_{\bf k}$, {\it i.e.} an intraband transition.
In the above, we defined the currents on horizontal and vertical
$O$-$Cu$-$O$-links, living on $Cu$-sites. We could alternatively have
defined the currents on $Cu$-$Cu$-links, both in a three-band model
and in a one-band model. This definition would be relevant to the
study of $d$-density waves \cite{AM1988,DDW}. However, one would
expect a different dissipation term in that case, due to the finite
modulation vector of the ordered currents.
Note also that the $\omega_\nu^2$-terms in $S_{\rm{Q}}$ in Eq. (\ref{eff_theory}), 
equivalently the inertial terms in Eqs. (\ref{CIM_1}) and (\ref{4-state}), 
are of multiband origin.
\par
The quartic terms in $S_{\rm{C}}$ that would emerge from the above treatment 
are of the type $\alpha^{ijlm}_{{\bf r}_1 {\bf r}_2 {\bf r}_3 {\bf r}_4}
J^{i}_{{\bf r}_1} J^{j}_{{\bf r}_2} J^{l}_{{\bf r}_3} J^{m}_{{\bf r}_4}$. Note 
that for $i=j=x$, $l=m=y$, these terms include an Ashkin-Teller type of four-spin 
interaction, used in Ref. \onlinecite{Aji_Varma} to argue that the Ising type of 
singularity in specific heat would be quenched. $S_{\rm{C}}$ in 
Eqs. (\ref{CIM_1},\ref{4-state}) differs from the model of
Ref. \onlinecite{Aji_Varma} in several respects. However, a direct comparison is 
difficult, as it is not clear what physical quantitites the fields in 
Ref. \onlinecite{Aji_Varma} represent. Firstly, the Ising-exchange coupling 
terms in Eqs. (\ref{CIM_1},\ref{4-state}) are anisotropic, possibly highly 
anisotropic, due to the bond-character of the Ising variables. Moreover, the term 
$\tilde{\alpha}^{xy}_{{\bf r},{\bf r}'} J^{x}_{{\bf r}} J^{y}_{{\bf r}'}$ 
in Eqs. (\ref{CIM_1},\ref{4-state}) is absent in Ref. \onlinecite{Aji_Varma}. 
While this term may be perturbatively irrelevant, it is far from clear that 
$\tilde{\alpha}_{xy}$ is actually small. In addition, there also seems to be 
a discrepancy between the dissipation kernel $|\omega_\nu|/|{\bf q}|$ derived 
here and the one employed in Ref. \onlinecite{Aji_Varma}. 
\par
We expect our model to be generically useful in describing thermal and
quantum critical fluctuations of directed particle-hole bond variables
in fermionic lattice models.

\indent \textit{Acknowledgements}. This work was supported by the  
Research Council of Norway Grants No. 158518/431 and No. 158547/431 
(NANOMAT), and Grant No. 167498/V30 (STORFORSK). The hospitality 
of the Center for Advanced Study at The Norwegian Academy of Science 
and Letters is acknowledged, as well as useful discussions with 
C. M. Varma and Z. Tesanovic.


\begin{thebibliography}{99} 


\bibitem{Varma} C. M. Varma, Phys. Rev. B {\bf 73}, 155113 (2006). 

\bibitem{MFL_Varma} C. M. Varma, P. B. Littlewood, S. Schmitt-Rink,
E. Abrahams, and A. E. Ruckenstein, Phys. Rev. Lett., {\bf 63}, 1996 (1989).

\bibitem{Aji_Varma}  V. Aji and C. M. Varma, Phys. Rev. Lett. {\bf 99}, 
067003 (2007).  
 
\bibitem{Greiter_Thomale} M. Greiter and R. Thomale, Phys. Rev. Lett., {\bf 99}, 027005 (2007). 

\bibitem{Fauque_Mook} B. Fauque {\it et al}, Phys. Rev. Lett. {\bf 96}, 197001 (2006);
H. A. Mook {\it et al}, Talk at Aspen Center for Physics, August 2007. 

\bibitem{AM1988} I. Affleck and J. B. Marston, Phys. Rev. B {\bf 37}, 3774 (1988).

\bibitem{DDW} S. Chakravarty, R. B. Laughlin, D. K. Morr, and C. Nayak,
Phys. Rev. B {\bf 63}, 094503 (2001).

\bibitem{HS} R. L. Stratonovich, Dokl. Akad. Nauk SSSR  {\bf 2}, 1097 (1957);
J. Hubbard, Phys. Rev. Lett., {\bf 3}, 77 (1959). 

\bibitem{Lee_Choi} H. C. Lee and H.-Y. Choi, Phys. Rev. B {\bf 64},
  094508 (2001).

\bibitem{footnote_Belitz} Note that this procedure \cite{Hertz} of
  integrating out the fermions and expanding the logarithm in the case
  of a $|{\bf q}|=0$ order parameter and gapless fermions in general
  leads to singular coefficients and might not be very suitable for
  renormalization group analysis. See D. Belitz {\it et al},
  Rev. Mod. Phys. {\bf 77}, 579 (2005).

\bibitem{Hertz} J. A. Hertz, Phys. Rev B {\bf 14}, 1165 (1976).


\bibitem{footnote_2}
The scaling function for the susceptibility is 
$\chi({\bf q},\omega) = (Z/T^{(2-\eta)/z}) 
\Phi_{\pm} \left(   \frac{(cq)^z}{T}, \frac{\omega}{T}  \right)$, where
$T$ is temperature, $Z$ is related to critical amplitudes, $z$ is dynamical
critical exponent, and $\eta$ is the anomalous scaling dimension of the 
relevant fields.  For $z = 3$, the ${\bf q}$-dependence of this quantity is
weak compared to the $\omega/T$-dependence at low $({\bf q},\omega)$ . 


\bibitem{Caldeira_Leggett} A. O. Caldeira and A. J. Leggett, Ann. Phys. (N.Y.)
{\bf 149}, 374 (1984).

\bibitem{footnote_anisotropy} Anisotropy is 
generic to bond-variables, which, unlike site-variables, have 
{\it directionality}. See also A. Melikyan and Z. Tesanovic,  Phys. 
Rev. B {\bf 74}, 214511 (2005).

\bibitem{Chakravarty_2005} See also P. Werner, K. V{\"o}lker, M. Troyer, 
and S. Chakravarty, Phys. Rev. Lett., {\bf 94}, 047201 (2005). The dissipative
part of the action in this (spatially extended) transverse field Ising-chain  
is non-local only in time, but local in space, due to the local character of
the coupling between the heat-bath oscillators and the Ising spins.
 
\bibitem{Onsager} L. Onsager, Phys. Rev. {\bf 65}, 117 (1944).


\end{thebibliography}
\end{document}